%% file: paper.tex
\documentclass[10pt,conference]{IEEEtran}
\usepackage[utf8]{inputenc}

\usepackage{soul}
\usepackage{xcolor}
\usepackage{xspace}
\usepackage[hyphens]{url}
\usepackage{graphicx}
\usepackage{listings,multicol}
\usepackage[caption=false]{subfig}  
\usepackage[export]{adjustbox}
\usepackage{amsmath}
\usepackage{amssymb}
\usepackage{textcomp}
\usepackage{breakurl} 
\usepackage{listings}
\usepackage{cleveref}
\usepackage{algorithm}
\usepackage[noend]{algpseudocode}
\usepackage{courier}
\usepackage{balance}

\newcommand{\iac}{\textsc{IaC}\xspace}
\newcommand{\iic}{\textsc{IiC}\xspace}

\title{Infrastructure in Code: \\ Towards Developer-Friendly Cloud Applications}

\author{
 \IEEEauthorblockN{Vladislav Tankov,\IEEEauthorrefmark{3}\IEEEauthorrefmark{1}\IEEEauthorrefmark{2}  Dmitriy Valchuk,\IEEEauthorrefmark{1}\IEEEauthorrefmark{4}  Yaroslav Golubev,\IEEEauthorrefmark{2}  Timofey Bryksin\IEEEauthorrefmark{2}\IEEEauthorrefmark{3}}
    \IEEEauthorblockA{\IEEEauthorrefmark{1}\textit{JetBrains},  \IEEEauthorrefmark{2}\textit{JetBrains Research},  \IEEEauthorrefmark{3}\textit{Higher School of Economics}, \IEEEauthorrefmark{4}\textit{ITMO University}}
    \IEEEauthorblockA{\{vladislav.tankov, valchuk.dmitriy, yaroslav.golubev, timofey.bryksin\}@jetbrains.com}
}

\begin{document}

\maketitle

\begin{abstract}
    The popularity of cloud technologies has led to the development of a new type of applications that specifically target cloud environments. Such applications require a lot of cloud infrastructure to run, which brought about the \textsc{Infrastructure as Code} approach, where the infrastructure is also coded using a separate language in parallel to the main application. In this paper, we propose a new concept of \textsc{Infrastructure in Code}, where the infrastructure is deduced from the application code itself, without the need for separate specifications. We describe this concept, discuss existing solutions that can be classified as \textsc{Infrastructure in Code} and their limitations, and then present our own framework called Kotless --- an extendable cloud-agnostic serverless framework for Kotlin that supports two cloud providers, three DSLs, and two runtimes. Finally, we showcase the usefulness of Kotless by demonstrating its efficiency in migrating an existing application to a serverless environment.  
\end{abstract}

\input{sections/01-introduction}
\input{sections/02-infrastructure-in-code}
\input{sections/03-existing-implementations}
\input{sections/04-kotless}
\input{sections/05-evaluation}
\input{sections/06-conclusion}

\balance

\bibliographystyle{ieeetran}
\bibliography{IEEEabrv,paper}

\end{document}

%% file: sections/01-introduction.tex
\section{Introduction}\label{sec:introduction}

In the last several years, \textit{cloud computing} has been getting more and more popular with developers and enthusiasts alike~\cite{intro:gartner-forecast}.
Modern \textit{cloud-native applications}~\cite{intro:cloud-native} that are being developed specifically for cloud provide better performance and scalability, while being more cost effective and fault tolerant~\cite{intro:cloud-native-pluses}. This is especially true for \textit{serverless} applications~\cite{existing:serverless-gains-popularity}. 
Cloud-native applications differ structurally from traditional web applications:
their complex and distributed nature means that their development is trickier~\cite{intro:cloud-native-is-hard}. Developers need to not only create a new application that uses some cloud API but also create an \textit{infrastructure} for it. Basically, the way an application is created is twofold: it requires both  writing code and deploying the accompanying infrastructure. 

One of the popular solutions to this problem is the \textsc{Infrastructure as Code (IaC)}~\cite{intro:iac-definition} approach. Within this technique, the developer creates special definition files that declare infrastructure alongside the rest of the code and uses these files to deploy the application into the cloud platform. However, this approach also has its drawbacks: deploying a simple static website of about 1,500 lines of code to Amazon Web Services (AWS) could require one to write more than 600 lines of additional \iac code in a separate domain-specific language, as well as to know the granular specifics of this particular cloud platform.

In this paper, we present a novel approach to the infrastructure declaration that makes the development of cloud-native applications a lot easier and does not require the developer to have expert-level knowledge about the particular cloud platform. We call this approach \textsc{Infrastructure in Code (IiC)}, which implies, contrary to the \textsc{Infrastructure as Code}, that it does not require one to create separate infrastructure declaration files and instead deduces the infrastructure right from the code itself. In the paper, we describe our vision of the concept and its key characteristics, as well as discuss existing solutions and their drawbacks.

We also showcase the \iic approach in our tool called \textit{Kotless}~\cite{kotless}, a serverless framework for Kotlin. This tool parses developers' Kotlin code to create the necessary serverless infrastructure automatically. We presented the prototype of Kotless in our previous work~\cite{intro:kotless-paper}. Since then, the prototype grew into a full-fledged framework that now supports two different cloud providers (AWS and Microsoft Azure), three DSLs (Ktor, Spring Boot, and our own Kotless DSL), and two runtimes (JVM and GraalVM binaries), with more than 800 stars on GitHub and a constantly growing community. We describe how Kotless operates, how it implements the \iic features, and show how it drastically reduces the complexity of developing serverless applications in comparison to the \iac approach.
Our case study shows how a web application can be moved to a serverless format using Kotless in several simple actions, reducing the cost of hosting by 80\%. 

%% file: sections/02-infrastructure-in-code.tex
\section{Infrastructure in Code}\label{sec:infrastructure-in-code}

We define \textsc{Infrastructure in Code} as a way of managing and provisioning infrastructure through automatic inference of infrastructural requirements from the application code. 
For example, a developer may use a \textit{@Get("/get")} annotation in the code, declaring that this code is an HTTP endpoint responding to GET requests. The \iic tool should be able to understand this and automatically provision an HTTP server that would be able to serve as an HTTP endpoint.

The overall pipeline of developing an application with the \iic tool is as follows:

\begin{enumerate}
    \item The developer writes an application using predefined statements that are interpreted as infrastructural requirements by the tool.
    \item Once the application is finished, the developer triggers its deployment to the cloud.
    \item The \iic tool analyzes the application to infer all the necessary infrastructure.
    \item The \iic tool generates all the required infrastructure.
    \item The application is deployed to the created infrastructure.
\end{enumerate}

Therefore, for this approach to work, an \iic framework should provide the following two main components:
\begin{itemize}
    \item \textit{Development API} that is used during the development of an application and helps to express infrastructural requirements;
    \item \textit{Cloud Implementation} --- a set of specific cloud services that would be generated in response to the infrastructural requirements. 
\end{itemize}

\subsubsection{Development API}\label{sec:infrastructure-in-code:development}
Overall, Development API defines the way developers express infrastructural requirements in code. \iic tools can be divided into two different types depending on the Development API: 
\begin{itemize}
    \item tools that introduce their own Domain Specific Languages (DSLs) to declare infrastructural requirements in code;
    \item tools that are able to analyze existing applications written with existing frameworks and translate their API into infrastructural requirements automatically. 
\end{itemize}

While both approaches make sense, only the second one makes it possible to migrate existing applications to automatically generated infrastructure. As noted by Baldini et al.~\cite{paradigm:serverless-current-trends}, the inability to automatically convert existing applications into cloud-native (serverless, in this case) is one of the biggest challenges for cloud adoption as a whole. Thus, we believe that an \iic tool following the second approach would be much more useful in practice. 

\subsubsection{Cloud Implementation}

Cloud implementation is an array of services of a cloud provider that are deployed to meet infrastructural requirements of an application. However, we believe that the definition of \iic should not define what specific cloud implementation should be chosen. For example, there are different computing models available within modern cloud platforms: the code can be executed inside long-living virtual machines or using short-living serverless functions, and both choices have advantages and disadvantages~\cite{paradigm:serverless-economics}.
Moreover, there are different cloud providers to support. 
We believe that each \iic tool should be implementing infrastructure in a way that is suitable for a particular purpose. For instance, there should be \iic tools that deploy serverless applications and those that create Kubernetes deployments. Such a diversity makes it possible for developers to choose the tool that suits their needs best.

\subsubsection{Limitations}

The \iic approach has its limitations, since not all kinds of infrastructure can be or should be inferred from the code. For example, it is rather difficult to get a schema of a NoSQL database from the application code, and even impossible to do so if the schema is being built in runtime. At the same time, an \iic tool would be simply unable to handle applications with an arbitrary logic. 

\subsubsection{Interoperability}\label{sec:infrastructure-in-code:interoperability}

Because of these limitations, we do not consider \iic tools to be an instant replacement for the \iac approach. Instead, we believe that both approaches can work together, and the natural way of creating new applications should be to create applications fast via the \iic approach and, if necessary, migrate to \iac later. 

Such an interoperability can be achieved by using an \iac tool under the hood of the given \iic tool:
instead of using the cloud platform's API directly and doing all the routines of infrastructure generation, an \iic tool can just generate \iac code and use it to deploy the infrastructure. This solves two problems at once: we use reliable tools for deploying infrastructure and support interoperability with an \iac instrument. Users can even use \iac tools with \iic tools in one project --- part of the infrastructural code will simply be generated during deployment.

%% file: sections/03-existing-implementations.tex
\section{Existing \iic implementations}

There exist tools that can be considered implementations of the \iic approach. Most of them are implemented for the serverless computing model, which is understandable given the growing popularity of serverless computations~\cite{existing:serverless-gains-popularity}. 

\textbf{AWS Chalice}~\cite{chalice} is an \iic tool for Python developed by Amazon Web Services (AWS). The tool uses its own DSL to define HTTP APIs and event handlers, and makes it pretty easy to create simple serverless applications. However, AWS Chalice has certain drawbacks.
It introduces its own DSL to define even the most basic HTTP handlers, which makes it impossible to automatically migrate to it from existing Python Web frameworks, and only supports the AWS cloud platform.

\textbf{Zappa}~\cite{zappa} is an \iic tool for Python that infers infrastructural requirements from Django API. 
Similar to Chalice, this tool only supports AWS. Unfortunately, under the hood, Zappa just wraps the entire application into a container and publishes it to the AWS API Gateway.
Because of this, if a developer wishes to customize their application somehow or integrate it with other cloud services, they would need to use Zappa's own \iac dialect to define permissions and integrations.

\textbf{Osiris}~\cite{osiris} is an \iic tool for Kotlin with its own DSL to define HTTP API events. It is serverless-oriented and only supports AWS. Osiris does not interact with AWS services directly, like other tools do, instead, it generates CloudFormation~\cite{cloudformation} definitions and uses them to deploy infrastructure to AWS, allowing for the interoperability.

All of the listed tools have certain disadvantages. First of all, all of them target only one cloud service provider, AWS. Secondly, all the tools support only one specific DSL --- either their own custom one or API of some specific web framework. Finally, two of the three tools interact with the cloud platform directly instead of using an \iac tool, which hampers interoperability.

%% file: sections/04-kotless.tex
\section{Kotless}

We present our own implementation of the \iic approach that addresses the issues of the existing tools. Kotless~\cite{kotless} is a Kotlin \iic tool that uses the serverless computation model to execute applications. Its main features are:
\begin{itemize}
    \item \textbf{cloud-agnostic}: it supports AWS and Microsoft Azure, and could be extended to other could service providers;
    \item \textbf{DSL-agnostic}: it has its own DSL, but also supports two popular Web frameworks, Ktor~\cite{ktor} and Spring~\cite{spring};
    \item \textbf{runtime-agnostic}: it runs on JVM, but also compiles into binary via GraalVM;
    \item \textbf{interoperable}: it integrates well with Terraform, one of the most popular \iac tools.
\end{itemize}

\begin{figure}[htbp]
\centering
    \includegraphics[width=\columnwidth]{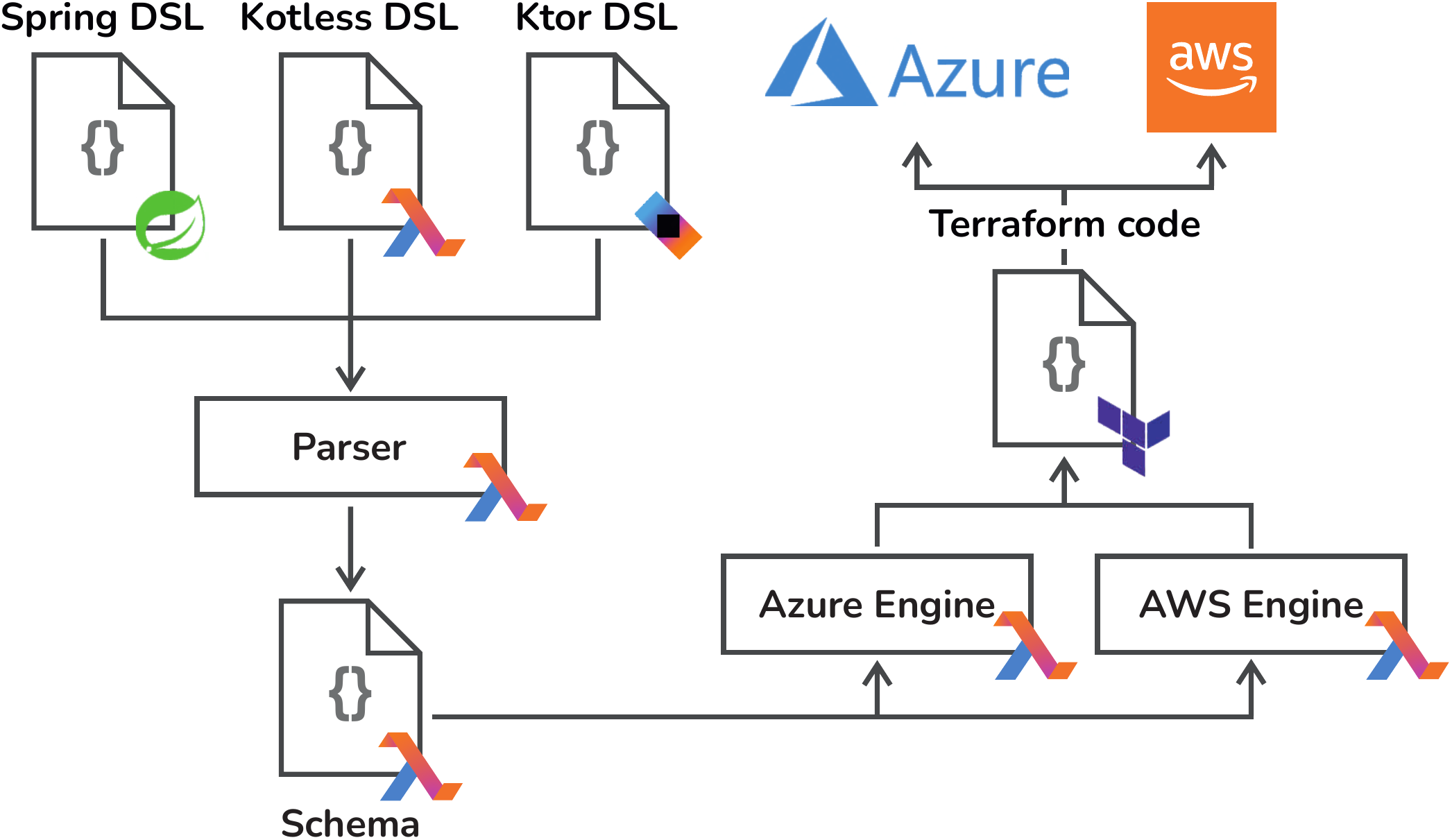}
    \centering
    \caption{The overall architecture of Kotless.}
    \vspace{-0.3cm}
    \label{fig:kotless-architecture}
\end{figure}

The pipeline of Kotless refines the overall \iic pipeline and is presented in Figure~\ref{fig:kotless-architecture}:

\begin{enumerate}
    \item The developer is writing code with one of the supported DSLs (see~\Cref{sec:kotless:parser}).
    \item The Parser module starts the Kotlin compiler and, during the analysis phase, gets infrastructural requirements via the adapter for the specific DSL.
    \item The infrastructural requirements are translated into the cloud-agnostic Schema.
    \item The Engine module runs the generator for a specific cloud platform and generates the necessary \iac code.
    \item The application is deployed via the \iac tool.
\end{enumerate}

Let us now describe each module in more detail.

\subsubsection{Cloud-agnostic schema}\label{sec:kotless:schema}

The cloud-agnostic schema is one of the key points of the Kotless architecture. It decouples the entire tool into two isolated components (analysis and generation) and makes it possible to introduce new DSLs and support new cloud platforms independently of each other without changing the existing code.

It consists of two main parts: \textit{general schema}, which expresses infrastructural requirements for different cloud platforms in a uniform way, and \textit{cloud-dependant extensions}, which allows to add events specific to particular cloud platforms (for example, S3 notifications).

The general schema introduces three main entities:
\begin{itemize}
    \item \textit{lambda}, which is a dynamic handler for events;
    \item \textit{static resource}, which is a static file that may be used in a response;
    \item \textit{permission}, which describes permissions associated with a specific \textit{lambda}.
\end{itemize}

For these entities, we introduce HTTP API definitions:
\begin{itemize}
    \item \textit{dynamic route}, which is a definition of an HTTP endpoint that should be processed by a lambda;
    \item \textit{static route}, which is a definition of an HTTP endpoint by request to which a static resource should be returned.
\end{itemize}

Of course, modern cloud platforms support events other than HTTP requests for serverless applications. The schema can be extended with custom events, for example, we added the \textit{Scheduled} event to define scheduled jobs within the cloud platform.
To support specific events of particular cloud platforms, we implemented interoperability with Terraform, an \iac tool. 
The developer can define a lambda and just subscribe it to some cloud-specific event with a piece of Terraform code. For example, this can be used to create a certain resource and then share it across different applications that are implemented in different architectures or on different clouds.

\subsubsection{Parser module}\label{sec:kotless:parser}

The Parser module is a platform for code analysis and a set of extensions that perform the inference of infrastructural requirements for specific DSLs.  
In this paper, we refer to the APIs of supported frameworks (annotations of Spring, extension functions of Ktor, etc.) also as DSLs, since from the perspective of \iic tools they are used to define infrastructural requirements in code.

Internally, the Kotless Parser module is implemented as a plugin to the Kotlin compiler. 
This gives us a great deal of flexibility: Kotless is able to not only get the information about annotations, but also to retrieve functions, types, and even perform compile-time calculations if needed. 
To make the implementation of the DSL analysis easier, Kotless provides a platform with extension points.
Currently, Kotless supports three DSLs: Ktor Web Framework API, Spring Boot API, and our own Kotless DSL. Their support was implemented via a common analysis platform and reusable analysis logic.
The comparison of these three DSLs is available online~\cite{dsl_comparison}.

\subsubsection{Engine module}\label{sec:kotless:engine}

Finally, to perform the actual deployment of the application to the cloud platform, we use the Engine module. As mentioned in~\Cref{sec:infrastructure-in-code:interoperability}, we opt for supporting interoperability between our \iic tool and popular \iac tools, so we do not implement the integration of Kotless with each cloud platform, but instead simply generate the \iac code. For that purpose, we chose Terraform---a well-known mature tool that  
supports almost all existing cloud platforms via a common language, HashiCorp Configuration Language (HCL). 
This language is typed and supports variables, links, string templates, etc. Each cloud platform has its own pack of so called \textit{resources} that are basically types in HCL and that are defined in HCL Schemas. 
To provide a convenient way of working with HCL code, we created a library called Terraform.kt~\cite{terraform_kt} that generates type-safe Kotlin DSL from HCL schemas and packs it into libraries. With Terraform.kt, writing generators of Terraform code in Kotlin is almost as simple as writing the Terraform code itself. Terraform.kt is also fully open-source.

The generation infrastructure consists of a set of generators. Each generator awaits for some Schema element as input and can depend on some other generator's output, thus, forming a dependency graph. 
During the generation process, the platform runs a visitor through the Kotless Schema object and executes all the generators that are able to run in the current context. Their outputs are added into the context, so other generators may run as soon as their dependencies are satisfied. When nothing else is able to run, the generation process stops and all the Terraform.kt entities are serialized into Terraform code.

Based on this infrastructure, we implemented support for AWS and Microsoft Azure. Currently, the support of Google Cloud and Yandex.Cloud is in process (the support of Yandex.Cloud is being added by Yandex developers themselves). 

In the case of AWS, we decided to use API Gateway as the main provider of HTTP endpoints. It supports the integration of static and dynamic routes and is fully serverless, which means that users do not have to pay for the application if it is not being currently used. To run lambdas, we used AWS Lambda, and we chose S3 to host static resources, since it is a reliable storage of static files, fully integrated with API Gateway. Finally, as a trigger for scheduled events, we used CloudWatch Events and their integration with AWS Lambda.

In the case of Microsoft Azure, we considered two HTTP endpoint providers: Azure Functions Proxy and API Management. While the latter is more powerful and supports more integrations, both were enough to implement static and dynamic routes. Since API Management was too expensive and slow to deploy, we chose Azure Functions Proxy. We decided to use Azure Functions to run lambdas and Azure Storage Blob to host static files. The scheduled trigger was implemented via Azure Functions Time Trigger.

\subsubsection{Running in the cloud}

Finally, the application should somehow integrate with the cloud platform API to work after being deployed. Each cloud platform provides a number of interfaces that applications should implement to be able to run and respond to events. To conform to this, Kotless implements so-called \textit{adapters}. Each adapter is specific to the particular cloud platform and is basically a special library that translates requests from a cloud platform into a necessary Web framework (Spring, Ktor, or any other) and translates back responses. Such adapters are implemented for each supported DSL and also for each supported cloud platform. 

An important aspect of running an application in the cloud is the specific runtime that the application uses. Kotlin is a multi-platform language that can be compiled into Java Virtual Machine (JVM) bytecode, JavaScript, and native binaries (via Kotlin/Native or GraalVM). 
Kotless supports two main runtimes for Kotlin: JVM (for all cloud platforms and DSLs) and GraalVM binaries (currently, only for AWS and Ktor). 

Supporting GraalVM is a unique Kotless feature that speeds up JVM-based serverless applications by orders of magnitude.
It also does not have a problem of a cold start~\cite{paradigm:serverless-current-trends} that is specific to the JVM runtime.
The drawback of GraalVM is that it requires special attention to the used Java APIs. 
For example, because of this limitation, the GraalVM runtime is currently supported only by Ktor DSL that does not heavily use reflection, in contrast to Spring and Kotless DSL.

%% file: sections/05-evaluation.tex
\section{Case study}\label{sec:evaluation}

To demonstrate the usefulness of the \iic approach and Kotless, let us consider the Play KotlinLang project~\cite{kotlinlang}. This is a Kotlin playground with a fully-featured built-in IDE: it includes a Kotlin compiler, syntax highlighting, and code completion, as well as virtual machines able to perform compilation of code and secure its execution. 

Overall, Play KotlinLang has an average workload of 2.23 requests per second, with a daily maximum of 17 requests per second. In total, it processes 5.8 million requests per month. However, the number of requests is not distributed evenly: we observe clear spikes during mornings and evenings. Because of that, with the old infrastructure that was based on permanent computing instances provided by the cloud platform, we had to over-provision those instances just to be sure the infrastructure would be able to handle these peak loads. This setup cost us about \$250 per month.

Such unevenly used application is a good candidate for moving to serverless. The only problem for us was the fact that Play KotlinLang was written using Spring Web Framework, and it was not clear how to migrate it into a serverless environment. The application is rather large (7,075 lines of Kotlin code) and too complex to simply deploy it as one serverless function: it has a lot of dynamic and static resources that belong to the front-end application.
Thus, we decided to use Kotless to perform the automatic migration to the serverless environment. All we had to do was to add one new class~\cite{class} that would serve as an entry point for the serverless platform and set up credentials to AWS. Kotless automatically introspected the entire code base and generated 989 lines of Terraform code. Then, the application was deployed to the cloud platform~\cite{kotlinlang_serverless}.

Overall, the process of migrating to serverless took about 15 minutes, while writing almost a thousand lines of Terraform code would take several hours in the best case scenario, including debugging via repeated deployments. The resulting Kotless-based version of the service costs us about \$50 per month, resulting in an 80\% saving rate. This example shows that Kotless is able to solve the problem of legacy applications migration~\cite{paradigm:serverless-current-trends} and facilitates the decrease in hosting costs for such applications.

%% file: sections/06-conclusion.tex
\section{Conclusion}\label{sec:conclusion}

In this paper, we present the concept of \textsc{Infrastructure in Code} --- a way of writing cloud-native applications, where the infrastructure code is generated automatically from the application code itself, in contrast to popular \textsc{Infrastructure as Code} tools that require developers to write separate deployment code. We discuss the main features and requirements of the \iic approach and describe existing tools supporting it.

To overcome the drawbacks of existing solutions, we present and describe Kotless --- a framework for creating serverless applications in Kotlin. Kotless can be extended in many directions: currently, it supports two different clouds (AWS and Microsoft Azure), three different DSLs, and two runtimes. Kotless was used to develop several applications, it saves developers from writing infrastructural code that could be twice as large as the main application. Kotless is also used in production at JetBrains, in other open-source projects, and has a growing community. We hope that it can be useful to practitioners all over the world. Also, while Kotless itself focuses on Kotlin, we would be happy if its assets were used by future developers to introduce brand new \iic tools for other languages, for example, Java or Python.

%% file: paper.bbl
\begin{thebibliography}{10}
\providecommand{\url}[1]{#1}
\csname url@samestyle\endcsname
\providecommand{\newblock}{\relax}
\providecommand{\bibinfo}[2]{#2}
\providecommand{\BIBentrySTDinterwordspacing}{\spaceskip=0pt\relax}
\providecommand{\BIBentryALTinterwordstretchfactor}{4}
\providecommand{\BIBentryALTinterwordspacing}{\spaceskip=\fontdimen2\font plus
\BIBentryALTinterwordstretchfactor\fontdimen3\font minus
  \fontdimen4\font\relax}
\providecommand{\BIBforeignlanguage}[2]{{%
\expandafter\ifx\csname l@#1\endcsname\relax
\typeout{** WARNING: IEEEtran.bst: No hyphenation pattern has been}%
\typeout{** loaded for the language `#1'. Using the pattern for}%
\typeout{** the default language instead.}%
\else
\language=\csname l@#1\endcsname
\fi
#2}}
\providecommand{\BIBdecl}{\relax}
\BIBdecl

\bibitem{intro:gartner-forecast}
``Gartner cloud forcast,''
  \url{https://www.gartner.com/en/newsroom/press-releases/2020-11-17-gartner-forecasts-worldwide-public-cloud-end-user-spending-to-grow-18-percent-in-2021},
  2020, accessed on 01.08.2021.

\bibitem{intro:cloud-native}
D.~Gannon, R.~Barga, and N.~Sundaresan, ``Cloud-native applications,''
  \emph{IEEE Cloud Computing}, vol.~4, no.~5, pp. 16--21, 2017.

\bibitem{intro:cloud-native-pluses}
D.~S. Linthicum, ``Cloud-native applications and cloud migration: The good, the
  bad, and the points between,'' \emph{IEEE Cloud Computing}, vol.~4, no.~5,
  pp. 12--14, 2017.

\bibitem{existing:serverless-gains-popularity}
P.~Castro, V.~Ishakian, V.~Muthusamy, and A.~Slominski, ``The rise of
  serverless computing,'' \emph{Communications of the ACM}, vol.~62, no.~12,
  pp. 44--54, 2019.

\bibitem{intro:cloud-native-is-hard}
C.~N.~C. Foundation, ``Top 7 challenges to becoming cloud native | cloud native
  computing foundation,''
  \url{https://www.cncf.io/blog/2020/09/15/top-7-challenges-to-becoming-cloud-native/},
  2020, accessed on 01.08.2021.

\bibitem{intro:iac-definition}
M.~{Artac}, T.~{Borovssak}, E.~{Di Nitto}, M.~{Guerriero}, and D.~A.
  {Tamburri}, ``Devops: Introducing infrastructure-as-code,'' in \emph{2017
  IEEE/ACM 39th International Conference on Software Engineering Companion
  (ICSE-C)}, 2017, pp. 497--498.

\bibitem{kotless}
``Kotless: A serverless framework for kotlin,''
  \url{https://github.com/JetBrains/kotless/}, accessed on 01.08.2021.

\bibitem{intro:kotless-paper}
V.~{Tankov}, Y.~{Golubev}, and T.~{Bryksin}, ``Kotless: A serverless framework
  for kotlin,'' in \emph{2019 34th IEEE/ACM International Conference on
  Automated Software Engineering (ASE)}, 2019, pp. 1110--1113.

\bibitem{paradigm:serverless-current-trends}
\BIBentryALTinterwordspacing
I.~Baldini, P.~C. Castro, K.~S. Chang, P.~Cheng, S.~J. Fink, V.~Ishakian,
  N.~Mitchell, V.~Muthusamy, R.~M. Rabbah, A.~Slominski, and P.~Suter,
  ``Serverless computing: Current trends and open problems,'' \emph{CoRR}, vol.
  abs/1706.03178, 2017. [Online]. Available:
  \url{http://arxiv.org/abs/1706.03178}
\BIBentrySTDinterwordspacing

\bibitem{paradigm:serverless-economics}
\BIBentryALTinterwordspacing
G.~Adzic and R.~Chatley, ``Serverless computing: Economic and architectural
  impact,'' in \emph{Proceedings of the 2017 11th Joint Meeting on Foundations
  of Software Engineering}, ser. ESEC/FSE 2017.\hskip 1em plus 0.5em minus
  0.4em\relax New York, NY, USA: Association for Computing Machinery, 2017, p.
  884–889. [Online]. Available: \url{https://doi.org/10.1145/3106237.3117767}
\BIBentrySTDinterwordspacing

\bibitem{chalice}
``Aws chalice: Python serverless microframework for aws,''
  \url{https://github.com/aws/chalice/}, accessed on 01.08.2021.

\bibitem{zappa}
``Zappa: Serverless python,'' \url{https://github.com/zappa/Zappa/}, accessed
  on 01.08.2021.

\bibitem{osiris}
``Osiris: Simple serverless web apps in kotlin,''
  \url{https://github.com/cjkent/osiris/}, accessed on 01.08.2021.

\bibitem{cloudformation}
``Cloudformation: An infrastructure as code instrument developed by aws,''
  \url{https://aws.amazon.com/cloudformation/}, accessed on 01.08.2021.

\bibitem{ktor}
``Ktor: A framework to build connected applications,'' \url{https://ktor.io/},
  accessed on 01.08.2021.

\bibitem{spring}
``Spring: Java framework,'' \url{https://spring.io/}, accessed on 01.08.2021.

\bibitem{dsl_comparison}
``Dsl comparison,'' \url{https://zenodo.org/record/4934743}, accessed on
  01.08.2021.

\bibitem{terraform_kt}
``Terraform.kt: a library for working with hcl via kotlin,''
  \url{https://github.com/TanVD/terraform.kt/}, accessed on 01.08.2021.

\bibitem{kotlinlang}
``Play kotlinlang,'' \url{https://play.kotless.io/}, accessed on 01.08.2021.

\bibitem{class}
``Additional class for play kotlinlang,''
  \url{https://github.com/TanVD/kotlin-compiler-server/blob/kotless/src/main/kotlin/com/compiler/server/CompilerApplication.kt},
  accessed on 01.08.2021.

\bibitem{kotlinlang_serverless}
``Kotless-based version of play kotlinlang,''
  \url{https://github.com/TanVD/kotlin-compiler-server/tree/kotless/}, accessed
  on 01.08.2021.

\end{thebibliography}
